\documentstyle[twoside,fleqn,espcrc2,epsfig]{article}

\newcommand{\AmS}{{\protect\the\textfont2
  A\kern-.1667em\lower.5ex\hbox{M}\kern-.125emS}}

\def    \np     #1#2#3{{\it Nucl. Phys.} {\bf #1}(19#2)#3}
\def    \pl     #1#2#3{{\it Phys. Lett.} {\bf #1}(19#2)#3}

\def    \pr     #1#2#3{{\it Phys. Rev.} {\bf #1}(19#2)#3}

\def    \hepph  #1 {{\tt hep-ph/#1}}
\def    \hepex  #1 {{\tt hep-ex/#1}}
\def\beq{\begin{equation}}
\def\eeq{\end{equation}}
\def\beqn{\begin{eqnarray}}
\def\eeqn{\end{eqnarray}}
\def\sss{\scriptscriptstyle}

\def\ptg{p_{{\sss T}\gamma}}
\def\etag{\eta_\gamma}
\def\xtg{x_{\sss T}^\gamma}

\def\ep{\epsilon}

\def\rightrightarrows{\rlap{\lower 2.5 pt \hbox{$\mathchar\rightarrow$}} 
                      \raise 1pt \hbox {$\mathchar\rightarrow$}}
\def\rightleftarrows{\rlap{\lower 2.5 pt \hbox{$\mathchar\leftarrow$}} 
                     \raise 1pt \hbox {$\mathchar\rightarrow$}}

\begin{document}

\twocolumn[
\vskip -3cm
~

\flushright{
        \begin{minipage}{2.9cm}
        CERN-TH/99-280\hfill \\
        hep-ph/9909360\hfill \\
        \end{minipage}        }

\vskip 5.0em
\flushleft{\Large A note on the production of photons at RHIC}~\footnotemark
\vskip\baselineskip
\flushleft{Stefano Frixione$^{\rm a}$}
\vskip\baselineskip
\flushleft{$^{\rm a}$CERN, TH division, Geneva, Switzerland}

\begin{center}
\begin{minipage}{160mm}
                \parindent=10pt
{\small 
I study the production of prompt photons in polarized hadronic collisions, 
considering different isolation prescriptions. In particular, I focus on 
the problem of the measurability of the polarized gluon density in the 
proton through isolated-photon data.
}
                \par
                \end{minipage}
                \vskip 2pc \par
\end{center}
]
\footnotetext{Talk given at QCD99, 7-13 July 1999, Montpellier, F.}


\section{Isolated-photon cross sections}

The study of the production of prompt photons in hadronic collisions is a 
very promising tool in at least two respects: first, it can be helpful
in the investigation of the underlying parton dynamics and second, it
is the (almost) only way to directly measure the gluon density of 
the nucleons at intermediate and large $x$'s. Although the cross
section for prompt-photon production is sizeably smaller than that
for jet or single-inclusive hadron production, the signal of the former
process is much cleaner than that of the latter processes. This is
basically due to the fact that the photon can couple directly only
to quarks, and not to gluons. This implies that, at the leading
order in perturbative QCD, only two partonic processes, namely
$qg\to\gamma q$ and $q\bar{q}\to\gamma g$, contribute to prompt-photon
cross section, as opposed to the larger number of processes 
in the case of jet production.

Unfortunately, the cleanliness of the prompt-photon signal is
limited by the fact that, as is well known, photons can also be
produced through a fragmentation process. In such a process, a
quark or a gluon, produced in a pure-QCD reaction, fragments into
a photon plus a number of hadrons. 
Furthermore, prompt photons have a dominant background due
to the production of neutral pions, with the subsequent decay
$\pi^0\to\gamma\gamma$. The problem of background rejection is
very effectively solved by requiring the photon to be isolated
from energetic hadron tracks in a `small' region surrounding
the photon itself. The nature of this region depends upon the type 
of the particles that initiate the scattering process: in $e^+e^-$ 
collisions, it is a cone of fixed aperture drawn around the
photon axis, while in hadronic collisions it is a subset of
the pseudorapidity-azimuthal angle plane, whose centre is given
by the pseudorapidity and azimuthal angle of the photon. The
difference in the definitions is due to the fact that, in the
case of hadronic collisions, one necessarily needs a prescription
invariant under longitudinal boosts. 
Apart from greatly reducing the background-to-noise ratio, the
isolation condition also diminishes the contribution to the cross
section coming from the fragmentation mechanism, relative to the
contribution of the direct mechanism, in which the photon 
participates in the hard scattering. This is due to the fact that,
in the fragmentation process, the photon and the companion hadrons, generated
by the fragmentation of the same parton, are usually close to each
other (fragmentation is essentially collinear), while in the direct
process the photon is usually well separated from other hadrons
(at the leading order, the photon and the recoiling hadron are
back-to-back in the transverse-momentum plane, because of the necessity
of momentum conservation).

In order to sensibly compare the data with the theoretical predictions,
it is essential that the very same isolation prescription is applied
both in the experimental analysis and in the computation of the cross
sections. From the theoretical point of view, this poses additional
problems with respect to the case of fully-inclusive photon production,
since the calculations are technically more involved, especially for
the fragmentation component. In general, when going beyond leading 
order in QCD, the direct and the fragmentation contributions are 
not separately well defined, both being divergent order by order
in perturbation theory. In the direct part, the divergence arises
from the collinear splitting of a quark into a photon and a quark.
Such a singularity is only cancelled by means of the bare
parton-to-photon fragmentation functions, which enter the fragmentation
contribution. Thus, even for very tight isolation prescriptions, the 
isolated-photon cross section depends upon the fragmentation mechanism.
However, an isolation definition has been proposed~\cite{iso98},
which is such that the cross section gets contribution only from 
the direct mechanism, and still is well defined at all orders in 
perturbation theory. This is seemingly incompatible with what
stated above concerning the cancellation of the QED collinear
singularity. However, with the definition of ref.~\cite{iso98},
the direct contribution is free of quark-photon collinear
singularities. Loosely speaking, this is achieved by requiring 
the energy of a parton to be smaller and smaller the closer the
parton is to the photon, until eventually only zero-energy partons
are allowed exactly collinear to the photon. In this way, the
vanishing energy of the quark damps the quark-photon collinear
singularity. More precisely, the isolation prescription of 
ref.~\cite{iso98} is given as follows: drawing a cone of half-angle 
$R_0$ around the photon axis in the $\eta -\phi$ plane (isolation cone), 
and denoting by $E_{T,had}(R)$ the total amount of transverse hadronic 
energy inside a cone of half-angle $R$, the photon is isolated if the 
following inequality is satisfied:
\beq
E_{T,had}(R)\le \ep_\gamma p_{{\sss T}\gamma} {\cal Y}(R),
\label{isoB}
\eeq
for all $R\le R_0$. Here, $p_{{\sss T}\gamma}$ is the transverse momentum of 
the photon. The function ${\cal Y}$ can be rather freely chosen, provided
that it vanishes fast enough for $R\to 0$. A sensible choice is the
following:
\beq
{\cal Y}(R)=\left(\frac{1-\cos R}{1-\cos R_0}\right)^n,
\;\;\;\;n=1.
\label{isfun}
\eeq
Notice that this isolation prescription is rather similar to
the ordinary cone prescription~\cite{cone}, which is obtained
by imposing:
\beq
E_{T,had}(R_0)\le \ep_c p_{{\sss T}\gamma}.
\label{isoA}
\eeq
Indeed, eq.~(\ref{isoA}) can be recovered from eqs.~(\ref{isoB})
and~(\ref{isfun}) by setting $n=0$ and $\ep_\gamma=\ep_c$ ($E_{T,had}(R)$
is by construction a function monotonically increasing with $R$).
The isolation prescriptions given in eqs.~(\ref{isoB}) and~(\ref{isoA})
have been proven to be infrared safe. More details can be found
in refs.~\cite{iso98,CFP}

\section{Photon production in polarized $pp$ collisions}

The isolation prescriptions given above apply to any kind of
hadron-hadron scattering. With only minor modifications (the 
isolation cone is drawn in the physical 3-space, and the transverse
energy is substituted with the energy), they can also be used
in $e^+e^-$ collisions. In the following, I will present phenomenological
predictions relevant to polarized $pp$ collisions, in the energy
range of the BNL collider RHIC ($\sqrt{S}=$200--500~GeV). All 
the results given in this paper are accurate to next-to-leading
order in QCD. A much more detailed and thorough discussion can be 
found in ref.~\cite{fv}.

One of the main goals of the collider RHIC will be that of extracting
the polarized gluon density in the proton, $\Delta g$, from prompt-photon
data. This will be done by comparing the experimental results with
the cross sections computed at the highest possible order in perturbation
theory (which is at present the next-to-leading one). It is therefore 
crucial to understand if the theoretical predictions for isolated-photon
production are reliable. This is not trivial, since the isolation cuts
are effective only on the radiative-emission contribution to the 
cross section, while the virtual corrections are not affected by
them; this means that the cancellation of the infrared
divergencies is perturbed by the isolation, regardless
of the fact that the isolation prescription is formally infrared safe.
As is customary in the cases in which a next-to-next-to-leading order
computation is not available, the perturbative stability of the
results can only be studied by looking at dependence of the physical
observables upon the renormalization and factorization scales.
As was shown in ref.~\cite{fv}, isolated-photon cross sections
at RHIC appear to be under good perturbative control, both for
inclusive isolated-photon and for photon-plus-jet observables. This
conclusion holds for both the isolation prescriptions of eqs.~(\ref{isoB})
and~(\ref{isoA}). It turns out that, when varying the scales between 
half and twice their default values, the cross sections change at the most
by a factor of 15\% (10\%) at the leading (next-to-leading) order. It
is therefore clear that the inclusion of the radiative corrections
improves the stability of the results, thus implying a non-pathological
behaviour of the perturbative expansion. 

Having established that the predictions of perturbative QCD can
be sensibly used in a comparison with data, I now turn to the issue
of the dependence of the isolated-photon cross sections upon the
polarized parton densities, in order to answer the question of whether
the measurements performed at RHIC will be useful in order to pin
down the presently poorly known $\Delta g$. The best tool to this
end is the asymmetry cross section ${\cal A}$; in fact, many systematic 
uncertainties cancel out in the ratio of polarized and unpolarized cross 
sections. The measurability of a spin asymmetry for a given process, as far 
as statistics is concerned, is of course  determined by the counting rate. 
The quantity 
\beq
\left({\cal A}\right)_{min}=\frac{1}{P^2}
\frac{1}{\sqrt{2\sigma {\cal L}\ep}}
\label{minasy}
\eeq
can be regarded as the minimal asymmetry that can be detected experimentally
or, equivalently, as the expected statistical error of the measurement,
for a given integrated luminosity relevant to parallel or antiparallel
spins of the incoming particles, ${\cal L}$, beam polarizations $P$ and 
a detection efficiency $\ep\le 1$; $\sigma$ is the unpolarized cross section 
integrated over a certain range in the observable under study. 
Eq.~(\ref{minasy}) can be obtained starting from the requirement that
the asymmetry be larger than its statistical error. The simplest
way to do so is to compute the statistical error affecting the
quantity $(N^{\rightrightarrows}-N^{\rightleftarrows})/
(N^{\rightrightarrows}+N^{\rightleftarrows})$, where $N$ is the number of 
production events for a given spin configuration (of the incoming particles). 
This quantity coincides with the asymmetry in the hypothesis in which the 
integrated luminosities for different spin configurations are
equal, which I assume here. With a straightforward computation,
assuming the statistical error on $N$ to be $\sqrt{N}$, and 
neglecting a factor of the order of $\sqrt{1-{\cal A}^2}$,
one gets immediately eq.~(\ref{minasy}).

\begin{figure}[htb]
\vspace{9pt}
\epsfig{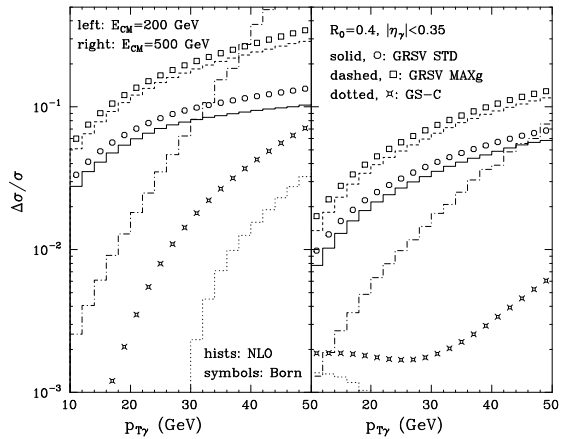}
\caption{Asymmetry as a function of $\ptg$, for $\sqrt{S}=200$~GeV
(left) and 500~GeV (right).}
\label{fig:ptasy}
\end{figure}
In fig.~\ref{fig:ptasy} I present the asymmetry as a function of the 
photon transverse momentum, for different centre-of-mass energies
of the colliding protons. A cut $|\etag |<0.35$ on the photon
pseudorapidity has been applied, and the isolation prescription of 
eq.~(\ref{isoB}) has been adopted. Both the next-to-leading order (histograms)
and the leading order (symbols) results are shown. The asymmetries have
been computed with three different parametrizations for the parton
densities. The GRSV STD set~\cite{GRSV} (which results in the solid
histograms) is the `best fit' (within the assumptions of the authors)
to the presently available data on polarized structure functions.
The GRSV MAXg~\cite{GRSV} and GS-C~\cite{GSC} sets (which result
in the dashed and in the dotted histograms respectively) have on the
other hand to be considered as the two most extreme choices compatible
with the structure function data, thus giving an estimate of the largest 
possible spread for the predictions of the asymmetry. Finally, the dot-dashed
histograms depict the minimally observable asymmetry as given in 
eq.~(\ref{minasy}), for ${\cal L}=100$~pb$^{-1}$, $P=1$ and $\ep=1$,
and with the unpolarized cross section integrated over a $\ptg$ bin
of width equal to 2~GeV. From the figure, we see that the shapes of 
the asymmetries obtained using the GRSV STD and GRSV MAXg sets are 
quite similar, but the difference in normalization is sizeable. 
On the other hand, the result obtained with GS-C looks completely 
different; the asymmetry turns negative over a certain
$\ptg$ range at the next-to-leading order. It is clear that,
if the parton densities are similar to those of the GS-C set,
it will be rather difficult to get a non-zero signal for the
asymmetry at RHIC. A more favourable situation can be found if
a larger pseudorapidity range is considered ($-1<\etag <2$).
The minimally observable asymmetry decreases by a factor that 
can be as large as 2, and the asymmetry obtained with GS-C sizeably
increases, becoming larger than $({\cal A})_{min}$ in the
low-$\ptg$ region.

It is instructive to compare the asymmetries at the two different 
centre-of-mass energies considered in fig.~\ref{fig:ptasy}. We can
observe that, as is well known, at smaller centre-of-mass energies 
the asymmetries are generally larger. However, when comparing 
the predicted asymmetries with the minimally observable 
asymmetry, it is clear that, at a fixed value of $\ptg$, and except 
for the first few $\ptg$ bins, the situation at $\sqrt{S}=500$~GeV
is more favourable than that at $\sqrt{S}=200$~GeV.
On the other hand, as far as the measurement of $\Delta g$ at a given $x$ is 
concerned, one should rather look at the asymmetries at fixed 
$\xtg=2 \ptg/\sqrt{S}$, since this corresponds to the value at which 
the parton densities are probed predominantly. 
Then, the quantity deciding about which energy is more favourable, is the 
minimally observable asymmetry at a given $\xtg$. For $\sqrt{S}=500$~GeV,
one finds a value of $\left({\cal A}_{\ptg}\right)_{min}$ larger than for 
the lower energy, making the higher-energy option appear less favourable. 
However, two points should be kept in mind here: firstly, in both plots
in fig.~\ref{fig:ptasy} the same value for the integrated luminosity
has been used, whereas in reality one anticipates a higher (by a factor 
of 2 to 3) luminosity for $\sqrt{S}=500$~GeV. Secondly, the lower
cut-off for $\ptg$ will certainly be the same for both energies, 
which means that at $\sqrt{S}=500$~GeV one can explore a region of
$\xtg$ that is inaccessible at $\sqrt{S}=200$~GeV. 

The same asymmetries as presented above have been computed adopting the
isolation prescription of eq.~(\ref{isoA}). Only negligible differences
were found in the case of the GRSV density sets, while some difference
can be observed, in the case of the GS-C set, in the central $\etag$
region, where the asymmetry obtained with eq.~(\ref{isoA}) is smaller
than that obtained with eq.~(\ref{isoB}). Also, asymmetries were computed,
for both isolation prescriptions, as a function of quantities defined
in terms of the photon and of the recoiling jet variables. In general, 
and as far as the statistics is concerned, inclusive-photon measurements 
seem to be somewhat more favourable than photon-plus-jet ones. More details
can be found in ref.~\cite{fv}.

\section*{Acknowledgements}
The results presented in this paper have been obtained in collaboration
with W.~Vogelsang, whom I warmly thank. The partial financial support
by the University of Montpellier is gratefully acknowledged.

\end{document}